\documentclass[aps,prl,superscriptaddress,showpacs,floatfix,twocolumn]{revtex4}

\usepackage{graphicx}

\begin{document}


\title{Jet Structure of Baryon Excess in Au+Au Collisions at 
$\sqrt{s_{NN}}=$200 GeV}

\newcommand{\abilene}{Abilene Christian University, Abilene, TX 79699, USA}
\newcommand{\acadsin}{Institute of Physics, Academia Sinica, Taipei 11529, Taiwan}
\newcommand{\banaras}{Department of Physics, Banaras Hindu University, Varanasi 221005, India}
\newcommand{\barc}{Bhabha Atomic Research Centre, Bombay 400 085, India}
\newcommand{\bnl}{Brookhaven National Laboratory, Upton, NY 11973-5000, USA}
\newcommand{\caucr}{University of California - Riverside, Riverside, CA 92521, USA}
\newcommand{\ciae}{China Institute of Atomic Energy (CIAE), Beijing, People's Republic of China}
\newcommand{\cns}{Center for Nuclear Study, Graduate School of Science, University of Tokyo, 7-3-1 Hongo, Bunkyo, Tokyo 113-0033, Japan}
\newcommand{\colorado}{University of Colorado, Boulder, CO 80309}
\newcommand{\columbia}{Columbia University, New York, NY 10027 and Nevis Laboratories, Irvington, NY 10533, USA}
\newcommand{\dapnia}{Dapnia, CEA Saclay, F-91191, Gif-sur-Yvette, France}
\newcommand{\debrecen}{Debrecen University, H-4010 Debrecen, Egyetem t{\'e}r 1, Hungary}
\newcommand{\elte}{ELTE, E{\"o}tv{\"o}s Lor{\'a}nd University, H - 1117 Budapest, P{\'a}zm{\'a}ny P. s. 1/A, Hungary}
\newcommand{\fsu}{Florida State University, Tallahassee, FL 32306, USA}
\newcommand{\gsu}{Georgia State University, Atlanta, GA 30303, USA}
\newcommand{\hiroshima}{Hiroshima University, Kagamiyama, Higashi-Hiroshima 739-8526, Japan}
\newcommand{\ihepprot}{Institute for High Energy Physics (IHEP), Protvino, Russia}
\newcommand{\illuiuc}{University of Illinois at Urbana-Champaign, Urbana, IL 61801}
\newcommand{\isu}{Iowa State University, Ames, IA 50011, USA}
\newcommand{\jinrdubna}{Joint Institute for Nuclear Research, 141980 Dubna, Moscow Region, Russia}
\newcommand{\kaeri}{KAERI, Cyclotron Application Laboratory, Seoul, South Korea}
\newcommand{\kangnung}{Kangnung National University, Kangnung 210-702, South Korea}
\newcommand{\kek}{KEK, High Energy Accelerator Research Organization, Tsukuba-shi, Ibaraki-ken 305-0801, Japan}
\newcommand{\kfki}{KFKI Research Institute for Particle and Nuclear Physics (RMKI), H-1525 Budapest 114, POBox 49, Hungary}
\newcommand{\korea}{Korea University, Seoul, 136-701, Korea}
\newcommand{\kurchatov}{Russian Research Center ``Kurchatov Institute", Moscow, Russia}
\newcommand{\kyoto}{Kyoto University, Kyoto 606, Japan}
\newcommand{\labllr}{Laboratoire Leprince-Ringuet, Ecole Polytechnique, CNRS-IN2P3, Route de Saclay, F-91128, Palaiseau, France}
\newcommand{\lawllnl}{Lawrence Livermore National Laboratory, Livermore, CA 94550, USA}
\newcommand{\losalamos}{Los Alamos National Laboratory, Los Alamos, NM 87545, USA}
\newcommand{\lpc}{LPC, Universit{\'e} Blaise Pascal, CNRS-IN2P3, Clermont-Fd, 63177 Aubiere Cedex, France}
\newcommand{\lund}{Department of Physics, Lund University, Box 118, SE-221 00 Lund, Sweden}
\newcommand{\muenster}{Institut f\"ur Kernphysik, University of Muenster, D-48149 Muenster, Germany}
\newcommand{\myongji}{Myongji University, Yongin, Kyonggido 449-728, Korea}
\newcommand{\nagasaki}{Nagasaki Institute of Applied Science, Nagasaki-shi, Nagasaki 851-0193, Japan}
\newcommand{\newmex}{University of New Mexico, Albuquerque, NM 87131, USA}
\newcommand{\nmsu}{New Mexico State University, Las Cruces, NM 88003, USA}
\newcommand{\ornl}{Oak Ridge National Laboratory, Oak Ridge, TN 37831, USA}
\newcommand{\orsay}{IPN-Orsay, Universite Paris Sud, CNRS-IN2P3, BP1, F-91406, Orsay, France}
\newcommand{\peking}{Peking University, Beijing, People's Republic of China}
\newcommand{\pnpi}{PNPI, Petersburg Nuclear Physics Institute, Gatchina, Russia}
\newcommand{\riken}{RIKEN (The Institute of Physical and Chemical Research), Wako, Saitama 351-0198, JAPAN}
\newcommand{\rikenrbrc}{RIKEN BNL Research Center, Brookhaven National Laboratory, Upton, NY 11973-5000, USA}
\newcommand{\saispbstu}{St. Petersburg State Technical University, St. Petersburg, Russia}
\newcommand{\saopaulo}{Universidade de S{\~a}o Paulo, Instituto de F\'{\i}sica, Caixa Postal 66318, S{\~a}o Paulo CEP05315-970, Brazil}
\newcommand{\seoulnat}{System Electronics Laboratory, Seoul National University, Seoul, South Korea}
\newcommand{\stonybrkc}{Chemistry Department, Stony Brook University, Stony Brook, SUNY, NY 11794-3400, USA}
\newcommand{\stonycrkp}{Department of Physics and Astronomy, Stony Brook University, SUNY, Stony Brook, NY 11794, USA}
\newcommand{\subatech}{SUBATECH (Ecole des Mines de Nantes, CNRS-IN2P3, Universit{\'e} de Nantes) BP 20722 - 44307, Nantes, France}
\newcommand{\tenn}{University of Tennessee, Knoxville, TN 37996, USA}
\newcommand{\titech}{Department of Physics, Tokyo Institute of Technology, Tokyo, 152-8551, Japan}
\newcommand{\tsukuba}{Institute of Physics, University of Tsukuba, Tsukuba, Ibaraki 305, Japan}
\newcommand{\vandy}{Vanderbilt University, Nashville, TN 37235, USA}
\newcommand{\waseda}{Waseda University, Advanced Research Institute for Science and Engineering, 17 Kikui-cho, Shinjuku-ku, Tokyo 162-0044, Japan}
\newcommand{\weizmann}{Weizmann Institute, Rehovot 76100, Israel}
\newcommand{\yonsei}{Yonsei University, IPAP, Seoul 120-749, Korea}
\affiliation{\abilene}
\affiliation{\acadsin}
\affiliation{\banaras}
\affiliation{\barc}
\affiliation{\bnl}
\affiliation{\caucr}
\affiliation{\ciae}
\affiliation{\cns}
\affiliation{\colorado}
\affiliation{\columbia}
\affiliation{\dapnia}
\affiliation{\debrecen}
\affiliation{\elte}
\affiliation{\fsu}
\affiliation{\gsu}
\affiliation{\hiroshima}
\affiliation{\ihepprot}
\affiliation{\illuiuc}
\affiliation{\isu}
\affiliation{\jinrdubna}
\affiliation{\kaeri}
\affiliation{\kangnung}
\affiliation{\kek}
\affiliation{\kfki}
\affiliation{\korea}
\affiliation{\kurchatov}
\affiliation{\kyoto}
\affiliation{\labllr}
\affiliation{\lawllnl}
\affiliation{\losalamos}
\affiliation{\lpc}
\affiliation{\lund}
\affiliation{\muenster}
\affiliation{\myongji}
\affiliation{\nagasaki}
\affiliation{\newmex}
\affiliation{\nmsu}
\affiliation{\ornl}
\affiliation{\orsay}
\affiliation{\peking}
\affiliation{\pnpi}
\affiliation{\riken}
\affiliation{\rikenrbrc}
\affiliation{\saispbstu}
\affiliation{\saopaulo}
\affiliation{\seoulnat}
\affiliation{\stonybrkc}
\affiliation{\stonycrkp}
\affiliation{\subatech}
\affiliation{\tenn}
\affiliation{\titech}
\affiliation{\tsukuba}
\affiliation{\vandy}
\affiliation{\waseda}
\affiliation{\weizmann}
\affiliation{\yonsei}
\author{S.S.~Adler}	\affiliation{\bnl}
\author{S.~Afanasiev}	\affiliation{\jinrdubna}
\author{C.~Aidala}	\affiliation{\bnl} \affiliation{\columbia}
\author{N.N.~Ajitanand}	\affiliation{\stonybrkc}
\author{Y.~Akiba}	\affiliation{\kek} \affiliation{\riken}
\author{A.~Al-Jamel}	\affiliation{\nmsu}
\author{J.~Alexander}	\affiliation{\stonybrkc}
\author{R.~Amirikas}	\affiliation{\fsu}
\author{K.~Aoki}	\affiliation{\kyoto}
\author{L.~Aphecetche}	\affiliation{\subatech}
% ...
\collaboration{PHENIX Collaboration} \noaffiliation

\date{\today}

%

\begin{abstract}
Two particle correlations between 
identified meson and baryon trigger particles with $2.5 < p_T < 4.0$ GeV/$c$
and lower $p_T$ charged hadrons have been measured at midrapidity by 
the PHENIX experiment at RHIC in p+p, d+Au and Au+Au collisions
at $\sqrt{s_{NN}}$=200 GeV.  The probability of
finding a hadron near in azimuthal angle to the trigger particle
is almost identical for leading mesons
and baryons for non-central Au+Au.
The yield for both trigger baryons and mesons is significantly higher in 
Au+Au than in p+p and d+Au, except for trigger baryons in central
collisions.
The baryon excess is likely to arise predominantly from hard
scattering processes.

\end{abstract}
\maketitle


A remarkable feature of relativistic heavy ion collisions is 
 greatly enhanced production of baryons and anti-baryons
relative to mesons. This enhancement over elementary
p+p collisions occurs at transverse momenta ($p_T$) 
of 2--5 GeV/$c$ $p_T$ \cite{ppg006,ppg015,STARlambda}. 
In this 
range, particle production shifts from 
soft processes (non-perturbative, low momentum transfer scattering)
to hard 
(high momentum transfer parton-parton scattering). Hard scattering is
followed by fragmentation of the scattered partons to jets of hadrons.
Baryon and anti-baryon production is suppressed in fragmentation.
Phenomenologically, this can be thought of as a large penalty for 
creating a diquark/anti-diquark pair for baryon formation vs. a
quark/anti-quark pair for meson formation. 

Since there is no sharp
separation of scales between hard and soft processes,
it is natural to ask which is responsible for the baryon excess in 
Au+Au collisions.
Hadron formation by recombination of boosted quarks from a collectively 
expanding source 
explains the observed baryon/meson
ratios \cite{HwaPRC,Fries,Ko,FriesQM}. The models include
recombination of 
quarks from a thermal
source, 
but the handling of hard partons varies significantly. Some allow
hard parton fragmentation only, with no recombination \cite{Fries}.
Others model the distribution of shower partons in a jet, and allow
soft and hard partons to coalesce \cite{Hwa}. These models should
produce different hadron-hadron correlations in the
recombination region, however quantitative predictions are not
available. 
Baryon production by recombination of purely thermal quarks
implies that the baryon excess is of soft origin, 
not from jet fragmentation. However, the
yield of baryons in this momentum range scales approximately
with the number of binary nucleon-nucleon collisions \cite{ppg015}, 
which is typical of hard processes.
Hadron production via recombination between jet
fragments and thermal quarks \cite{Hwa}
could preserve jet-like correlations among the final hadrons,
presuming that each hadron contains at least one quark arising
from a fragmenting hard scattered parton. As baryons contain
one quark more than mesons, baryon production may be more strongly 
enhanced by the availability of additional quarks. Observation of
such a mechanism would indicate modification of the jet
fragmentation process by the medium produced in Au+Au collisions.

To determine the role 
of jets in the production 
of intermediate $p_T$ protons, the PHENIX experiment at RHIC 
has measured
energetic hadronic partners near the baryons. These are
the additional fragmentation products from the same jet as the baryon. We
present first results on two particle correlations where the 
\emph{trigger} particle is an identified meson ($\pi$,$K$) or
baryon ($p$,$\bar{p})$ at 2.5 $< p_T< $  4.0 GeV/$c$. 
\emph{Associated} particles, i.e. lower $p_T$ 
charged hadrons, near in azimuthal angle  to the trigger
are counted.  Momentum cuts are chosen to avoid contamination by
resonance decays. The centrality and collision system
dependence of the associated particle yield per trigger is used
to help disentangle thermal quark recombination from jet fragmentation.
Trigger particles 
from recombination of boosted thermal quarks should
not have correlated partners beyond the expected correlation from
elliptic flow. However, if the source of the baryons
is indeed fragmentation of hard scattered partons, the
probability of finding a jet-like hadronic partner should
be comparable to that observed in p+p collisions.
We use p+p collisions without trigger identification to 
provide a comparison baseline.

%
Data presented here include collisions at $\sqrt{s_{NN}} = 200$~GeV
of Au+Au (24 million events),  d+Au (42 million events) and p+p 
(23 million events).
Charged particles are reconstructed in the central arms of
PHENIX using drift chambers, each with an azimuthal coverage
of $\pi/2$
and two layers of multi-wire proportional chambers with pad
readout (PC1, PC3)~\cite{nim_phenix}.  Pattern recognition
is based on a combinatorial Hough transform in the track bend
plane, with the polar angle determined by PC1 and 
the collision vertex along the beam direction~\cite{ppg006}.
Particle momenta are measured with a resolution
$\delta p/p = 0.7\% \oplus 1.0\%p~(\rm{GeV}/c)$ in Au+Au and
$\delta p/p = 0.7\% \oplus 1.1\%p~(\rm{GeV}/c)$ in d+Au and p+p.
The portion of the east
arm spectrometer containing the high resolution Time-Of-Flight(TOF) 
detector, which covers pseudo-rapidity $|\eta| < 0.35$ and
$\phi = \pi/4$ in azimuthal angle is used
for trigger particle identification.  
Beam Counters (BBC)~\cite{nim_phenix} provide
the global start; stop signals are from TOF scintillators
at a radial distance of 5.06 m.  The timing resolution is
$\sigma$ = 120 ps, which allows a 4$\sigma$ separation of mesons/baryons
up to $p_T \approx$ 4 GeV/$c$.  The Au+Au centrality determination
is described in Ref.~\cite{ppg26}.


Distributions of azimuthal angular difference, 
$\Delta\phi$, are constructed for trigger-partner pairs.
The combinatorial background is determined by constructing mixed
events
in two steps: 
the number of trigger and partner 
particles is determined by sampling the measured single
particle multiplicity distributions in the relevant momentum
and centrality 
ranges. Then the 3-momenta of particles in the mixed event
are determined by sampling
the measured trigger and partner momentum distributions.
To correct for the limited acceptance of the PHENIX
spectrometers, the real event $\Delta\phi$ distributions
are divided by $\Delta\phi$ distributions from the mixed events,
normalized to a constant angular aperture.
The shape of this distribution retains effects of the PHENIX azimuthal 
acceptance, but has no true correlations. The partner yield is  then
corrected for the reconstruction efficiency, detector aperture
and (for Au+Au only) detector occupancy~\cite{ppg23}. No extrapolation is
made to $|\eta| >$ 0.35.
Since d+Au and Au+Au collisions contain uncorrelated combinatorial background 
from other particles in the underlying event, the mixed event
partner yield per trigger, after the same efficiency correction,
 is subtracted using the normalization
determined by the convolution of the trigger and partner
single particle rates. 

Because mixing and subtraction is
done in finite size centrality bins, the background distribution
is biased toward the more central events within a bin, as
they produce more particles.
Consequently, the subtraction is corrected 
for the width of the centrality bins used for 
mixing: 5\% in Au+Au and minimum bias in d+Au. The width of 
trigger and partner number within a centrality
bin is determined from the 
measured centrality dependence of particle multiplicity in
the relevant momentum region and particle 
species~\cite{ppg23,ppg26,dAu_cent,dAu_id}. This width implies
larger fluctuations in the number of partners per trigger in mixed
events, so the mixed event partner yields are increased accordingly.
This correction modifies the background level by 
 $\approx$0.2\% in the most central
and $\approx$25\% in the most peripheral Au+Au collisions.

Elliptic flow causes an angular correlation in Au+Au  
unrelated to jet fragmentation, a background
to this measurement. The elliptic flow correlation is removed 
by modulating the azimuthally uniform combinatorial background by
$1+2v_{2}^{assoc}v_{2}^{trig}\cos(\Delta\phi)$, where
$v_{2}^{assoc}$ and $v_{2}^{trig}$ are the $v_2$ values
measured for the partner and trigger $p_T$ ranges, 
respectively~\cite{ppg22} where the reaction plane is 
measured by the BBC at $3 < \eta < 4$  minimizing
the influence of jets in the $v_2$ values.  Because the centrality binning
in this analysis is finer than in \cite{ppg22} the 
$p_T$ integrated centrality dependence is used to
interpolate $v_2$ for collisions more central than 20\%.

Systematic uncertainties in Au+Au and d+Au partner 
yields arise from uncertainties in the corrections for 
centrality bin width, systematic and statistical errors on 
$v_2$ ~\cite{ppg22} 
(Au+Au only), uncertainty in the background subtraction due to the event
mixing technique, and uncertainty in the detector occupancy 
correction.
The cross-contamination of mesons and protons is less than 5\%.
The error on the occupancy correction reaches a maximum of 5\% in the
most central Au+Au collisions. 
For most 
Au+Au bins, the dominant systematic uncertainty
on the partner yields is the uncertainty in $v_2$. This produces a
systematic error of approximately 0.01 partners per trigger baryon in
semi-central and central collisions; for trigger mesons, the corresponding
error is somewhat smaller. The event mixing uncertainty is approximately
comparable to the $v_2$ uncertainty in these bins. In peripheral Au+Au
collisions, the dominant systematic error in the partner yield arises
from the centrality bin width corrections and $v_2$ uncertainty. In
d+Au collisions, there is no $v_2$, and the partner yield uncertainty
is driven by the correction for centrality bias of mixed events.
In p+p collisions the systematic error is taken
to be the same size as the combinatorial background which is subtracted.
The total systematic errors are shown in Figure \ref{fig2}.

%

\begin{figure}[tb]
\includegraphics[width=1.0\linewidth]{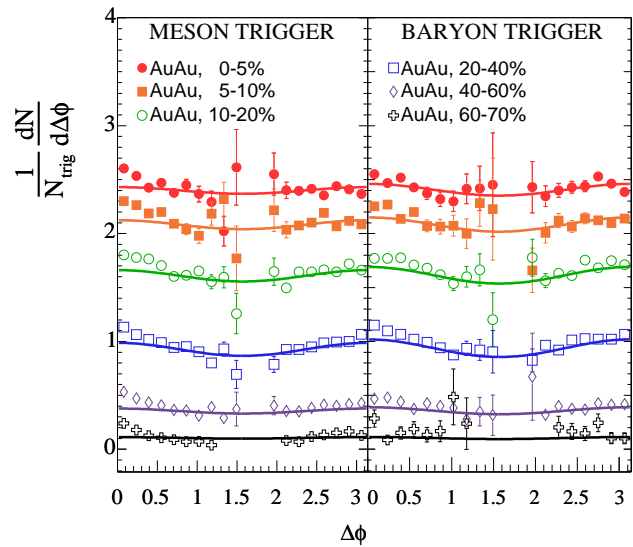}
\caption{\label{fig1} 
$\Delta\phi$ distributions for meson 
(left) and baryon (right) triggers with $2.5 < p_T < 4.0$ GeV/$c$  and associated 
charged hadrons with
$1.7 < p_T < 2.5$ GeV/$c$ for six centralities in Au+Au collisions.  The
solid lines indicate the calculated combinatorial background in the event
modulated by the measured elliptic flow.  
}
\end{figure} 

\begin{figure}[tb]
\includegraphics[width=1.0\linewidth]{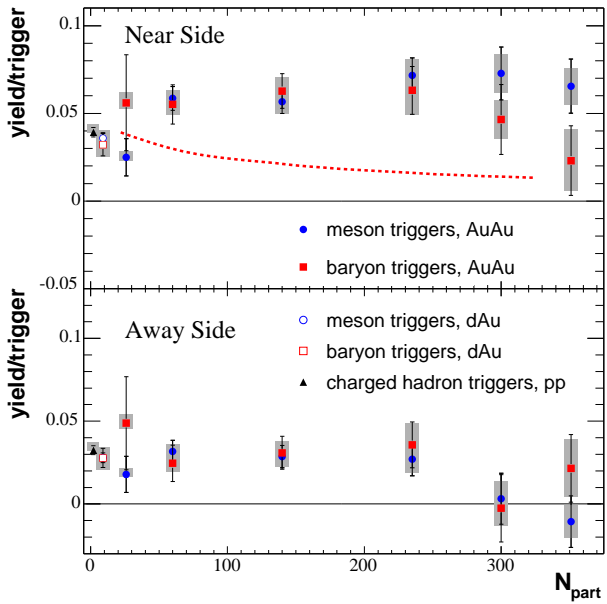}
\caption{\label{fig2} 
Yield per trigger for associated
charged hadrons between $1.7 < p_T < 2.5$ GeV/$c$ for the near- (top) and away-
(bottom) side jets.
The error bars are statistical errors and the grey boxes 
are systematic errors.
There is an additional 12\% error on the overall normalization
which 
moves all points together.  The dashed line (top)
represents an upper limit of the centrality dependence of the near-side
partner yield from thermal recombination (see text).
}
\end{figure} 

Figure~\ref{fig1} shows the 
$\Delta\phi$ distributions 
with trigger mesons (left) and baryons (right) measured
at midrapidity ($|\eta|<$0.35)  
in Au+Au collisions.
The jet-like correlations are clearly visible over 
the combinatorial background,
particularly at small relative angle inside the expected jet cone; 
the widths of the non-background distributions are consistent
with jet fragmentation. The background is subtracted,
and the number of associated partners per trigger integrated to
determine the conditional yield of partners. The near (far) side yield is
the integral over $0 < \Delta\phi < 0.94$ radians ($2.2 < \phi < \pi$ rad). 
The range is selected to include partners over as wide
a range as possible, while omitting the limited acceptance region around
$\Delta\phi = \pi/2$. 
However, the non-background associated partners
are observed in the angular range characteristic of jet fragmentation.
The near side jet width in p+p collisions
has been measured by PHENIX to be $\approx$ 0.25 rad
in a similar $p_T$ range~\cite{janqm}.

Figure~\ref{fig2} shows the conditional yield per trigger
of partner particles
in p+p, d+Au, and Au+Au collisions,
as a function of the number of participant nucleons.
The top panel shows partner yield at small relative angle,
from the same jet as the trigger hadron.
We observe an increase in partner yields 
in mid-central Au+Au compared to the d+Au and p+p collisions; 
this almost doubling of the near side partner yield 
suggests that the fragmentation is modified by the medium.
In Au+Au collisions, the near side yield per {\it meson}
trigger remains constant as a function of centrality, whereas the
near-side yield per {\it baryon} trigger decreases in the most 
central collisions as expected if a fraction of the baryons were
produced  by 
soft processes such as recombination of thermal quarks.
In d+Au collisions the near-side yields per trigger are the same for meson and
baryons triggers, and agree with results from p+p collisions generated
with PYTHIA~\cite{pythia}.

The dashed line in 
Figure~\ref{fig2} shows the 
expected centrality dependence of partners per baryon if all 
the ``extra" baryons \cite{ppg26} 
which increase $p/\pi$ over that in p+p collisions were to 
arise solely from soft processes. 
Baryons from thermal quark recombination
should have no jet-like partner hadrons and would dilute the
per-trigger conditional yield. 
Because this simple
estimate does not allow for meson production by recombination,
which must also occur along with baryon production,
it represents an upper limit to the centrality dependence of 
the jet partner yield from thermal recombination.  
The data clearly
disagree with both the centrality dependence and the absolute
yields  of this estimation,
indicating that the baryon excess has the same jet-like origin as
the mesons, except perhaps in the highest centrality bin.


The bottom panel of Figure 2 shows the
conditional yield of partners on the away side.
The partner yield in $2.2 < \Delta\phi < \pi$ rad
drops equally for both trigger baryons and mesons
going from p+p and d+Au to central Au+Au,
in agreement with the observed disappearance \cite{starb2b}
and/or broadening \cite{janqm} of the 
dijet
azimuthal correlations. It further supports the conclusion that
the baryons originate from the same jet-like mechanism as mesons.

\begin{figure}[ht]
\includegraphics[width=1.0\linewidth]{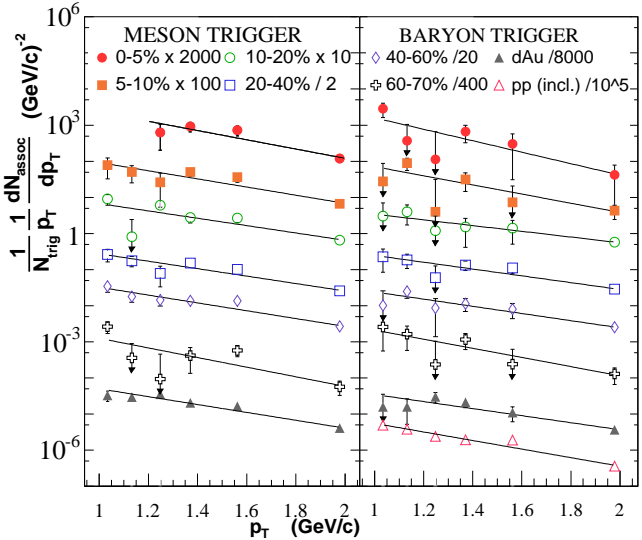}
\caption{\label{fig3}
$p_T$ spectra of the near side associated charged hadrons
corrected to the full jet yield 
for meson (left) and baryon (right) triggers at
$2.5 < p_T < 4.0$ GeV/$c$ and $|\eta|<$ 0.35 for
six centralities in Au+Au, d+Au and p+p collisions (non-identified
trigger).  Errors are
statistical only.  The curves are exponential fits.  Inverse
slope values are shown in Figure~\ref{fig4}.
}
\end{figure} 

\begin{figure}[ht]
\includegraphics[width=1.0\linewidth]{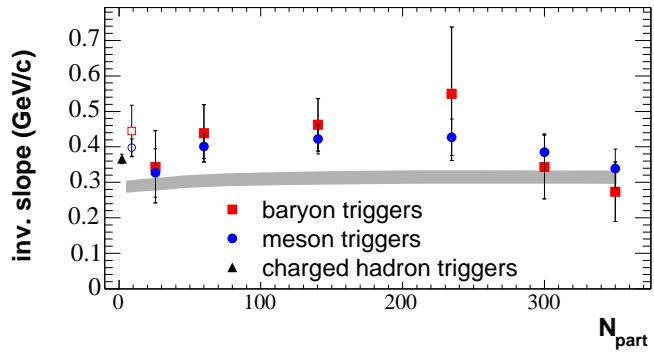}
\caption{\label{fig4}
Inverse slopes from the fits in Figure \ref{fig3}.
Solid (hollow) squares and circles are Au+Au (d+Au) collisons and
the triangle is p+p collisons.  The solid band indicates the 
slopes of inclusive
particle spectra in Au+Au collisions \cite{ppg23}.}
\end{figure}

Figure \ref{fig3} shows the $p_T$ spectra of particles associated 
on the near side with trigger mesons and baryons. 
The measured transverse momentum
of jet hadrons with respect to the inital parton 
direction, $\langle j_T \rangle$, 
is constant as a function of 
collision energy and $p_T$ ~\cite{ccor,janqm}.  Thus the 
angular size of jets increases as the partner $p_T$ decreases.
We used the PHENIX measurement of
$\langle |j_{Ty}|\rangle = 0.359 \pm 0.011$~GeV/$c$~\cite{janqm} to 
correct the near side conditional yield measured
in $\Delta\phi<$ 0.94 rad and the
PHENIX $\eta$ acceptance to the full jet yield. 
This correction assumes 
that jets are symmetric gaussians in both $\phi$ and $\eta$, 
and is 
required for the partner $p_T$ spectra because of the 
$p_T$ dependence of the jet width.
The conditional yields in Figure \ref{fig2} do not have
this additional correction as they are measured in a single
$p_T$ bin.  The partner spectra in Figure \ref{fig3} are fitted 
with exponentials, and the slopes are given in Figure \ref{fig4}
as a function of $N_{part}$. 
The jet partner slopes exceed those of inclusive hadrons in the
same $p_T$ range~\cite{ppg23},
except perhaps in the most central collisions.
The partner slopes are consistent between the collision systems and
trigger type, indicating a common jet-like source for both baryons and
mesons.

%

We have presented the first study of the jet structure of 
baryons ($p, \overline{p}$) and mesons ($\pi, K$)
at midrapidity in Au+Au collisions at $\sqrt{s_{NN}}$ = 200GeV, 
in the momentum region where baryon production is greatly
enhanced in central Au+Au.
Three observations indicate that mesons and baryons both arise predominantly
from hard processes in all but the most central Au+Au
collisions. First, 
baryons and mesons both have jet-like partner
particles. Second, 
there is no strong change of 
the slope of the  $p_T$ spectra of 
associated particles from  p+p to d+Au to Au+Au collisions,
and it is larger than that of inclusive hadrons. 
Finally, on the away side, the jet partner yield into a 0.94 radian opening
angle decreases in central collisions similarly for
trigger baryons and mesons.
The data are therefore inconsistent with a simple picture of baryon production 
at intermediate $p_T$ 
dominated by recombination of only thermal quarks.
On the 
trigger particle side, jets in
Au+Au collisions are modified compared to those in p+p. They are
richer in leading baryons and show enhanced probability for jet-like
partners, except for the most central collisions with trigger baryons.


We thank the staff of the Collider-Accelerator and Physics
Departments at BNL for their vital contributions.  We acknowledge
support from the Department of Energy and NSF (U.S.A.), 
MEXT and JSPS (Japan), CNPq and FAPESP (Brazil), NSFC (China), 
IN2P3/CNRS, CEA, and ARMINES (France), 
BMBF, DAAD, and AvH (Germany), 
OTKA (Hungary), DAE and DST (India), ISF (Israel), 
KRF and CHEP (Korea), RMIST, RAS, and RMAE (Russia), 
VR and KAW (Sweden), U.S. CRDF for the FSU, 
US-Hungarian NSF-OTKA-MTA, and US-Israel BSF.


\end{document}